\documentclass[aps,preprint]{revtex4}%
\usepackage{amsfonts}
\usepackage{amsmath}
\usepackage{amssymb}
\usepackage{graphicx}%
\begin{document}
\preprint{ }
\title{Reflection of electromagnetic waves from mixtures of plane gravitational and
scalar waves.}
\author{O. Gurtug}
\email{ozay.gurtug@emu.edu.tr}
\author{M. Halilsoy}
\email{mustafa.halilsoy@emu.edu.tr}
\author{O. Unver}
\email{ozlem.unver@emu.edu.tr}
\affiliation{Department of Physics, Eastern Mediterranean University, G.Magusa, North
Cyprus, Mersin 10 - Turkey}
\pacs{PACS number}

\begin{abstract}
We consider colliding wave packets consisting of hybrid mixtures of
electromagnetic, gravitational and scalar waves. Irrespective of the scalar
field, the electromagnetic wave still reflects from the gravitational wave.
Some reflection processes are given for different choice of packets in which
the Coulomb-like component $\Psi_{2}$ vanishes. Exact solution for multiple
reflection of an electromagnetic wave from successive impulsive gravitational
waves is obtained in a closed form. It is shown that a succesive sign flip in
the Maxwell spinor arises as a result of encountering with an impulsive train
( i.e. the Dirac's comb curvature) of gravitational waves. Such an observable
effect may be helpful in the detection of gravitational wave bursts.

\end{abstract}
\maketitle

\section{INTRODUCTION.}

Reflection of electromagnetic (em) waves in general relativity is totally
different from its counterpart process of classical electromagnetism. While in
the latter the incident em wave is partly reflected and transmitted, the
reflector remains unaffected. Corresponding process in general relativity is a
completely different process, as a manifestation of inherent non-linearity the
reflector wave pulse is modified as well. Thus exact solution of the full
Einstein - Maxwell (EM) equations respecting all boundary conditions becomes indispensable.

The problem of colliding em waves with gravitational waves has not yet been
solved in the general case but only in particular situations. The solution by
Griffiths corresponds to the most elementary one \cite{G1},\cite{G2}%
,\cite{G3}. Closely related to the subject are the solutions of colliding
impulsive light-like signals \cite{BH} and the light-like shells with
impulsive gravitational waves in scalar-tensor theory of gravity \cite{BR}.
The problem of reflection is interesting \ in the sense that it incorporates a
nonsymmetrical initial data in the collision process. The exact analytic
solution is obtained with an imposed condition that \ one of the metric
functions characterising the gravitational degrees of freedom depend only on
one of the null coordinates as it was in the pre-collision region. Significant
reduction in the field equations follows as a result of this condition.

In Griffiths's work, exact solutions for two different cases were given; one
corresponding to an impulsive, the other to a shock gravitational wave. The
main feature of the problem is that the em wave partly reflects from the
gravitational wave while the latter does not reflect from the em wave. As a
result of the nonlinear interaction a Coulomb-like (i.e. $\Psi_{2}$ in the
Newman - Penrose formalism) component develops apart from the modified wave
component $\Psi_{4}$. Another feature in this process is that in the
interaction region a space-like curvature singularity necessarily develops.

The purpose of this paper is to present new extensions to the same physical
process by employing the solution of Griffiths. One of the extensions is to
consider the collision and interaction of plane gravitational radiation and em
waves in the presence of a massless scalar field. Another extension is to
consider the non-linear interaction of a plane em wave with the succesion of
plane impulsive gravitational waves.

The main motivation to consider the scalar field in this problem is that, in
recent years there is a growing literature around the subject of scalar field
cosmology. With this in mind, it would be physically interesting to
investigate the effect of scalar field taking part in the reflection problem
of em waves. In addition to this, we are aiming to find a non-singular Cauchy
horizon (CH) forming reflection process. To incorporate scalar fields in this
direction, we construct CH forming Einstein-scalar and em-scalar plane wave
packets that propagate in the opposite directions in each of the incoming
regions as shown in Fig.1. The resulting space-time is as a result of the
collision of these wave packets.

The gravitational wave in our problem has a different structure compared to
the cases considered in the references \cite{G1},\cite{G2},\cite{G3}. In the
problem considered, the gravitational radiation is composed of a plane
impulsive part accompanied with shock waves. The obtained solution has
revealed in contrast to our expectation that, it does not yield a CH forming
reflection process. Our solution implies that the em wave still partly
reflects from the gravitational wave irrespective of the presence of the
scalar field. Following the collision, the resulting space-time has the
$\Psi_{4}$ and $\Psi_{2}$ components and they are both unbounded as the
focusing hypersurface is approached. General feature of the problem has not
been changed with the insertion of the scalar field, but we have shown that a
particular choice of the scalar field makes $\Psi_{2}=0$, with the only
non-zero component $\Psi_{4}$. This provides a unique example in which the
scalar field can rule the gravitational field.

Another extension is naturally provided by considering an em wave that
encounters a train of impulsive gravitational waves (i.e. the Dirac's comb).
The solution describes simply, how the em wave reflects from the successive
gravitational impulsive waves. In principle the problem is one pertaining to
multiple reflection but idealized choice of gravitational impulsive wave train
renders an exact solution possible. We show that the result is similar to the
single impulsive wave case, with the additional property that the reflecting
Maxwell component changes sign while crossing from one wave to the next. This
fact may play role in the identification of succesive impulsive gravitational
waves which remain still elusive to any detection technique. The whole
discussion can also be extended to a gravitational wave with non-aligned
polarization, however for convenience reasons we shall restrict ourselves in
this paper to the linear polarization.

Organization of the paper is as follows: In section 2, we construct CH forming
Einstein-scalar and em-scalar plane wave spacetimes. In section 3, we consider
the collision and subsequent interaction of these wave packets. A particular
scalar field that makes $\Psi_{2}=0$ is analysed in section 4. Reflection of
an em wave from a succession of impulsive waves follows in section 5 and we
complete the paper with our conclusion in section 6.

\section{Horizon Forming Einstein-scalar and Em-scalar Solutions.}

In order to obtain CH forming solutions with scalar fields we choose CH
forming pure em and pure gravitational solutions known beforehand. For the
pure gravitational solution we choose the solution found by Yurtsever
\cite{YU} and independently by Ferrari and Ibanez \cite{FE}. An interesting
property of this solution is that, its interaction region is isometric to the
part of the interior region of the Schwarzschild black hole.

As for the pure em case we employ the well-known Bell-Szekeres solution
\cite{BS}. The scalar field is added to both solutions by the $M$%
\textit{-shift} method which we had explained elsewhere \cite{GHS}. For
completeness we summarize this method briefly as follows: The general metric
for colliding plane waves is represented by \cite{G1},%

\begin{equation}
ds^{2}=2e^{-M}dudv-e^{-U}\left\{  \left[  e^{V}dx^{2}+e^{-V}dy^{2}\right]
\cosh W-2\sinh Wdxdy\right\}
\end{equation}

in which the metric functions depend at most on the null coordinates $u$ and
$v$. The basic field equations of colliding Einstein - Maxwell - scalar (CEMS)
system \ can be generated from the following Lagrangian density ( for parallel
polarization, $W=0$ ),%

\begin{equation}
L=e^{-U}\left\{  M_{u}U_{v}+M_{v}U_{u}+U_{u}U_{v}-V_{u}V_{v}-4\phi_{u}\phi
_{v}-2e^{U+V}A_{u}A_{v}\right\}  .
\end{equation}

Varying this Lagrangian yields the following CEMS field equations,%

\begin{align}
U_{uv}  &  =U_{u}U_{v},\\
2M_{uv}  &  =-U_{u}U_{v}+V_{u}V_{v}+4\phi_{u}\phi_{v},\\
2V_{uv}  &  =U_{v}V_{u}+U_{u}V_{v}-2\left(  \bar{\Phi_{0}}\Phi_{2}+\bar
{\Phi_{2}}\Phi_{0}\right)  ,\\
2\phi_{uv}  &  =U_{v}\phi_{u}+U_{u}\phi_{v},\\
2A_{uv}  &  =V_{v}A_{u}+V_{u}A_{v}.
\end{align}

The remaining two equations%

\begin{align}
2U_{uu}-U_{u}^{2}+2M_{u}U_{u}  &  =V_{u}^{2}+4\phi_{u}^{2}+4\Phi_{2}\bar
{\Phi_{2}},\nonumber\\
2U_{vv}-U_{v}^{2}+2M_{v}U_{v}  &  =V_{v}^{2}+4\phi_{v}^{2}+4\Phi_{0}\bar
{\Phi_{0}},
\end{align}

corresponding to $R_{uu}=-T_{uu}$ and $R_{vv}=-T_{vv}$ which do not follow
from the variational principle are used to integrate for the metric function
$M$. In these equations $\phi$ represents the massless scalar field, $\Phi
_{0}$ and $\Phi_{2}$ are the Newman-Penrose spinors for em fields defined by%

\begin{align}
\Phi_{2}  &  =F_{\mu\nu}\overline{m}^{\mu}n^{\nu}=-\frac{e^{\frac{U-V}{2}}%
}{\sqrt{2}}A_{u},\nonumber\\
\Phi_{0}  &  =F_{\mu\nu}l^{\mu}m^{\nu}=\frac{e^{\frac{U-V}{2}}}{\sqrt{2}}%
A_{v},\nonumber\\
F_{\mu\nu}  &  =\partial_{\mu}A_{\nu}-\partial_{\nu}A_{\mu},
\end{align}

where subscripts denote partial derivatives and overbar denotes the complex
conjugate. The non-zero Weyl and Ricci scalars are defined by%

\begin{align}
\Psi_{0}  &  =-C_{\mu\nu\rho\sigma}l^{\mu}m^{\nu}l^{\rho}m^{\sigma},\\
\Psi_{2}  &  =-C_{\mu\nu\rho\sigma}l^{\mu}m^{\nu}\overline{m}^{\rho}n^{\sigma
},\nonumber\\
\Psi_{4}  &  =-C_{\mu\nu\rho\sigma}n^{\mu}\overline{m}^{\nu}n^{\rho}%
\overline{m}^{\sigma},\nonumber\\
\Phi_{00}  &  =-\frac{1}{2}R_{\mu\nu}l^{\mu}l^{\nu},\nonumber\\
\Phi_{11}  &  =-\frac{1}{4}R_{\mu\nu}\left(  l^{\mu}n^{\nu}+m^{\mu}%
\overline{m}^{\nu}\right)  ,\nonumber\\
\Phi_{22}  &  =-\frac{1}{2}R_{\mu\nu}n^{\mu}n^{\nu},\nonumber\\
\Phi_{02}  &  =-\frac{1}{2}R_{\mu\nu}m^{\mu}m^{\nu},\nonumber\\
\Lambda &  =\frac{1}{24}R,\nonumber
\end{align}

where $C_{\mu\nu\rho\sigma}$ stands for the Weyl tensor, $R_{\mu\nu}$ the
Ricci tensor and $R$ is the Ricci scalar. Weyl scalars have the following
physical interpretation: The $\Psi_{0}$ and $\Psi_{4}$ terms represents the
transverse gravitational wave components in the $\ n^{\mu}$ and $l^{\mu}$
direction, respectively. The $\Psi_{2}$ component is known as the "Coulomb -
like" component that arises as a result of non-linear interaction. The Ricci
scalars \ $\Phi_{00}$, $\Phi_{11}$, $\Phi_{02}(=\overline{\Phi}_{20})$,
$\Phi_{22}$ and $\Lambda$\ \ on the other hand, corresponds to the matter
fields. In terms of these Ricci scalars, the energy-stress tensor is given by%

\begin{align}
4\pi T_{\mu\nu} &  =\Phi_{00}n_{\mu}n_{\nu}+\Phi_{22}l_{\mu}l_{\nu}+\Phi
_{02}\overline{m}_{\mu}\overline{m}_{\nu}+\Phi_{20}m_{\mu}m_{\nu}+\left(
\Phi_{11}+3\Lambda\right)  \left(  l_{\mu}n_{\nu}+n_{\mu}l_{\nu}\right)  \\
&  +\left(  \Phi_{11}-3\Lambda\right)  \left(  m_{\mu}\overline{m}_{\nu
}+\overline{m}_{\mu}m_{\nu}\right)  .\nonumber
\end{align}

We incorporate \ a scalar field $\phi(u,v)$ into this vacuum ( or
electrovacuum) metric through shifting the metric function $M$ in equations
(8)$,$ ( i.e. the $M$\textit{-shift, }for details see Ref. \cite{GHS} )%

\begin{equation}
M\rightarrow M+\Gamma
\end{equation}

where the scalar field is related to $\Gamma$ through the conditions%

\begin{align}
U_{u}\Gamma_{u}  &  =2\phi_{u}^{2},\\
U_{v}\Gamma_{v}  &  =2\phi_{v}^{2}.\nonumber
\end{align}

Stated otherwise, by shifting $M$ in accordance with (12) and through
identifications (13) the shift function $\Gamma$ accounts for the scalar field
in a consistent manner.\ The integrability condition imposes the massless
scalar field equation as a constraint condition,%

\begin{equation}
2\phi_{uv}-U_{u}\phi_{v}-U_{v}\phi_{u}=0.
\end{equation}

The metric function $U$ is already known from the vacuum (electrovacuum)
problem so that the crucial point is to choose a suitable scalar field that
will yield a CH instead of a curvature singularity. To achieve this part of
the problem it is often more convenient to pass to a prolate type of
coordinates ( $\tau,\sigma$ ) instead of the null coordinates ( $u,v$ ). For
the vacuum (electrovacuum) problem, the relation between the new coordinates (
$\tau,\sigma$ ) in terms of the null coordinates ( $u,v$ ) are defined by%

\begin{align}
\tau &  =\sin(au+bv),\\
\sigma &  =\sin(au-bv),\nonumber
\end{align}

where $a,b$ are constants. In the prolate coordinates the massless scalar
field equation (14) transforms into%

\begin{equation}
\left(  \Delta\phi_{\tau}\right)  _{\tau}-\left(  \delta\phi_{\sigma}\right)
_{\sigma}=0,
\end{equation}

where $\Delta=1-\tau^{2}$ and $\delta=1-\sigma^{2}$. The exact solutions to
above equation are already well-known \cite{G1}.

We observe that the choice%

\begin{equation}
\phi(\tau,\sigma)=a_{0}\tau\sigma\text{ \ \ \ \ \ \ \ \ \ \ \ \ \ \ \ }%
a_{0}=\text{constant\ ,\ }%
\end{equation}

solves the equation (16) and more importantly it belongs to a class of CH
forming solutions. Integrating the equations (13) for the metric function
$\Gamma$ in terms of the $\left(  \tau,\sigma\right)  $ coordinates yields%

\begin{equation}
\Gamma=a_{0}^{2}(\tau^{2}+\sigma^{2}\Delta).
\end{equation}

This function is well-behaved and finite as the focusing hypersurface is
approached. In the problem considered, the focusing hypersurface is given by%

\begin{equation}
e^{-U}=1-\sin^{2}au-\sin^{2}bv=\sqrt{\Delta\delta}=0,
\end{equation}

and in terms of the prolate coordinates, focusing hypersurface is defined for
$\tau=1$.

It should be noted that vanishing of the above equation at points $\sigma
=\pm1$ is not included in the interaction region. Those points represent the
null boundaries separating the interaction region from the incoming regions.

The metric that describes the collision of plane impulsive gravitational wave
accompanied with shock gravitational wave coupled with massless scalar field
is obtained for linear polarization ( $W=0$) case as follows

\begin{align}
e^{-U}  &  =\sqrt{\Delta\delta},\\
e^{-V}  &  =\sqrt{\frac{\delta}{\Delta}}\left(  1+\tau\right)  ^{2}%
,\nonumber\\
e^{-M}  &  =\left(  1+\tau\right)  ^{2}e^{-\Gamma}.\nonumber
\end{align}

The trace of the energy momentum tensor, \ $T$ ( $=R,$ the scalar curvature)
is obtained as%

\begin{equation}
T=R=\frac{4a_{0}^{2}e^{\Gamma}}{\left(  1+\tau\right)  ^{2}}\left(  \tau
^{2}-\sigma^{2}\right)  ,
\end{equation}

while the Ricci and Weyl scalars are given by%

\begin{align}
\Phi_{00}^{(0)}  &  =b^{2}\theta(v)a_{0}^{2}\sin^{2}\left(  2bv\right)  ,\\
\Phi_{22}^{(0)}  &  =a^{2}\theta(u)a_{0}^{2}\sin^{2}\left(  2au\right)  ,\\
\Phi_{11}^{(0)}  &  =-\frac{1}{2}ab\theta(v)\theta(u)a_{0}^{2}\sin\left(
2au\right)  \sin\left(  2bv\right)  ,\\
\Lambda^{(0)}  &  =\frac{1}{6}ab\theta(v)\theta(u)a_{0}^{2}\sin\left(
2au\right)  \sin\left(  2bv\right)  ,\\
\Psi_{2}^{(0)}  &  =ab\theta(v)\theta(u)\left[  \frac{1}{1+\tau}-a_{0}^{2}%
\sin\left(  2au\right)  \sin\left(  2bv\right)  \right]  ,\\
\Psi_{4}^{(0)}  &  =a\delta\left(  u\right)  -\frac{3a^{2}\theta(u)}{1+\tau
}+a_{0}^{2}a^{2}\theta(u)\sin\left(  2au\right)  \left[  2\cos\left(
au-bv\right)  -\sin\left(  2au\right)  \right]  ,\\
\Psi_{0}^{(0)}  &  =b\delta\left(  v\right)  -\frac{3b^{2}\theta(v)}{1+\tau
}+a_{0}^{2}b^{2}\theta(v)\sin\left(  2bv\right)  \left[  2\cos\left(
au-bv\right)  -\sin\left(  2bv\right)  \right]  ,
\end{align}

in which the scale invariant quantities are defined by $Z^{(0)}=e^{-M}Z$. Note
that the metric function $\Gamma$, the scalar $R,$ and the Weyl and Ricci
scalars all remain finite as the focusing hypersurface $\tau\rightarrow1$ is
approached. This indicates that curvature singularity does not develop on the
focusing surface and hence, that surface is called a Cauchy-horizon surface.

Similarly, the metric that describes CH forming em-scalar waves is%

\begin{align}
e^{-U}  &  =\sqrt{\Delta\delta},\\
e^{-V}  &  =\sqrt{\frac{\delta}{\Delta}},\nonumber\\
e^{-M}  &  =e^{-\Gamma},\nonumber
\end{align}

with the $\phi$ and $\Gamma$ once more as in equations (17) and (18),
respectively. The scale invariant Weyl and Ricci scalars in the present case are%

\begin{align}
\Psi_{2}^{\left(  0\right)  }  &  =-a_{0}^{2}ab\theta(u)\theta(v)\left(
\sin2au\right)  \left(  \sin2bv\right)  ,\\
\Psi_{4}^{\left(  0\right)  }  &  =a\delta(u)\theta(v)\tan\left(  bv\right)
+a_{0}^{2}a^{2}\theta(u)\theta(v)\left(  \sin2au\right)  \left(
\sin2bv\right)  ,\\
\Psi_{0}^{\left(  0\right)  }  &  =b\delta(v)\theta(u)\tan\left(  au\right)
+a_{0}^{2}b^{2}\theta(u)\theta(v)\left(  \sin2au\right)  \left(
\sin2bv\right)  ,\\
\Phi_{00}^{(0)}  &  =b^{2}\left[  1+a_{0}^{2}\sin^{2}\left(  2bv\right)
\right]  \theta(v),\\
\Phi_{22}^{(0)}  &  =a^{2}\left[  1+a_{0}^{2}\sin^{2}\left(  2au\right)
\right]  \theta(u),\\
\Phi_{11}^{(0)}  &  =-\frac{1}{2}a_{0}^{2}ab\sin\left(  2au\right)
\sin\left(  2av\right)  \theta(u)\theta(v),\\
\Lambda^{(0)}  &  =\frac{1}{6}a_{0}^{2}ab\theta(u)\theta(v)\sin\left(
2au\right)  \sin\left(  2av\right)  ,
\end{align}

which are all well-behaved functions in the interaction region ( $u>0,v>0$ ).
The distributional singularities are explicitly seen in (31) and (32) which
lie on the light cone. The non-zero scalar curvature invariant is given by%

\begin{align}
I  &  =\frac{1}{16}R_{\mu\nu\rho\sigma}R^{\mu\nu\rho\sigma}=2\left(  \Psi
_{0}\Psi_{4}+3\Psi_{2}^{2}\right) \\
&  =8a_{0}^{4}a^{2}b^{2}\theta(u)\theta(v)e^{2\Gamma}\sin^{2}\left(
2au\right)  \sin^{2}\left(  2bv\right)  ,\nonumber
\end{align}

which hosts no singularities.

\section{Reflection of em waves in the colliding Einstein-scalar and Em-scalar
waves.}

In the previous section, we have demonstrated, how the CH forming colliding
wave metrics can be constructed in the Einstein-scalar and Maxwell-scalar
theory. In this section, we shall consider the collision of the plane wave
packets that contain; plane impulsive gravitational waves accompanied with
shock gravitational waves coupled with massless scalar fields in one of the
incoming region II ( $u>0,v<0$ ), while in the other incoming region III (
$u<0,v>0$ ), we have plane em wave coupled with massless scalar fields (see Fig.1).

\subsection{Region II: $u>0,v<0$}

This region contains an Einstein-scalar plane wave which is represented by the
following metric,%

\begin{equation}
ds^{2}=2\left(  1+\sin au\right)  ^{2}e^{-\Gamma(u)}dudv-e^{-U(u)}\left\{
e^{V(u)}dx^{2}+e^{-V(u)}dy^{2}\right\}
\end{equation}

where%
\begin{align}
\Gamma(u)  &  =a_{0}^{2}\sin^{2}au\left(  1+\cos^{2}au\right)  ,\\
e^{-U(u)}  &  =\cos^{2}au,\nonumber\\
e^{-V(u)}  &  =\left(  1+\sin au\right)  ^{2},\nonumber
\end{align}

and the scalar field is%

\begin{equation}
\phi(u)=a_{0}\sin^{2}au.
\end{equation}

We note that the null coordinate$\ \ u$ is implied with a step function
$u\rightarrow u\theta(u)$ in all these expressions to assure that for $u<0$ we
have a flat space. This metric has the gravitational wave component%

\begin{equation}
\Psi_{4}^{(0)}=a\delta(u)-a^{2}\theta(u)\left[  \frac{3}{1+\sin au}-2a_{0}%
^{2}\cos au\sin2au\left(  1-\sin au\right)  \right]  ,
\end{equation}

which shows that it is a mixture of impulsive and shock waves.

\subsection{Region III: $u<0,v>0$}

This region contains an em-scalar plane wave which is given by the following metric,%

\begin{equation}
ds^{2}=2e^{-\Gamma(v)}dudv-\cos^{2}bv\left(  dx^{2}+dy^{2}\right)
\end{equation}

with%

\begin{align}
e^{-U(v)}  &  =\cos^{2}bv,\\
\Gamma(v)  &  =a_{0}^{2}\sin^{2}bv\left(  1+\cos^{2}bv\right)  ,\nonumber\\
\phi(v)  &  =-a_{0}\sin^{2}bv,\nonumber
\end{align}

and em potential $A_{\mu}=A\delta_{\mu}^{y}$ with%

\begin{equation}
A=\sqrt{2}\sin bv.
\end{equation}

The corresponding Maxwell field is then%

\begin{equation}
\Phi_{0}^{(0)}(v)=-b(\cos bv)e^{\frac{U}{2}}=-b,
\end{equation}

where $\Phi_{0}^{(0)}=e^{-M/2}\Phi_{0}$ $\left(  \text{and similarly }\Phi
_{2}^{(0)}=e^{-M/2}\Phi_{2}\right)  .$

\subsection{Region IV: $u>0,v>0$}

In order to obtain a solution in region IV which represents the collision of
these waves; we impose the same condition as Griffiths had introduced in the
references \cite{G1},\cite{G2},\cite{G3}. This condition is to take the metric
function $V=V(u)$. With this choice, the basic field equations ( 3-9 ) of CEMS
system simplifies to a great extend. The complete solution is given by%

\begin{align}
e^{-U} &  =\cos^{2}au+\cos^{2}bv-1,\\
e^{-V} &  =\left(  1+\sin au\right)  ^{2},\nonumber\\
e^{-M} &  =(\cos au)(\cos bv)e^{-\Gamma-V+U/2},\nonumber\\
\phi &  =a_{0}\left(  \sin^{2}au-\sin^{2}bv\right)  ,\nonumber\\
\Gamma &  =a_{0}^{2}\left(  \sin^{2}au+\sin^{2}bv\right)  \left(  \cos
^{2}au+\cos^{2}bv\right)  ,\nonumber\\
A_{y} &  =\sqrt{2}\sin bv\left(  1+\sin au\right)  ,\nonumber\\
\Phi_{0}^{(0)} &  =-b\theta(v)(\cos bv)e^{\frac{U}{2}},\nonumber\\
\Phi_{2}^{(0)} &  =\left(  \frac{V_{u}}{2}\right)  \left(  \sin bv\right)
e^{\frac{U}{2}},\nonumber
\end{align}

in which the null coordinates are as usual with the step functions,
$u=u\theta(u)$ and $v=v\theta(v)$.

The presence of $\Phi_{2}^{(0)}$ of the Maxwell field indicates that the em
field reflects from the gravitational field. The non-zero Weyl components are%

\begin{align}
\Psi_{2}^{(0)}  &  =-ab\theta(u)\theta(v)\left(  \sin2au\right)  \left(
\sin2bv\right)  \left(  a_{0}^{2}+\frac{e^{2U}}{4}\right)  ,\\
\Psi_{4}^{(0)}  &  =a\delta(u)-a^{2}\theta(u)\left(  1-\sin au\right)
\{\frac{1}{1+\sin au}+e^{U}\sin au\nonumber\\
&  -\frac{2}{3}a_{0}^{2}\sin au\left[  \cos2au+\cos2bv\right]  \},\nonumber
\end{align}

which indicate a curvature singularity on $e^{-U}=0$. The absence of the
$\Psi_{0}$ component reveals that the gravitational field does not reflect
from the em field. Our result verifies once more, in conform with the Mariot -
Robinson theorem \cite{G1} that reflection of em waves from gravitational
waves in a cosmology dominated by the particular scalar field considered above
remains unchanged. This result could not be anticipated a priori, unless
proved explicitly.

\section{A Particular scalar Field That Makes $\Psi_{2}=0.$}

The metric function $M$ is the one that determines the Weyl component
$\Psi_{2}$ since%

\begin{equation}
\Psi_{2}^{(0)}=\frac{1}{2}M_{uv}\text{.}%
\end{equation}

Is it possible to choose a scalar field that will lead to a separable $M(u,v)$
of the form $M(u,v)=M_{1}(u)+M_{2}(v)$ ?. This will enable us to set $\Psi
_{2}=0$, leaving behind only $\Psi_{4}$ for the gravitational field. By
inspecting (46) we see that the choice%

\begin{equation}
\phi(u,v)=\Gamma(u,v)=\frac{1}{2}U(u,v),
\end{equation}

does the trick and still it leads to an exact solution. In particular Eqs.
(13) and (14) are automatically satisfied by virtue of the field equations.
Thus, we can formulate our initial value problem of colliding
Einstein-Maxwell-scalar fields as follows

Region II:%

\begin{align}
\phi(u)  &  =\Gamma(u)=\frac{1}{2}U(u),\\
e^{-U(u)}  &  =\cos^{2}au,\nonumber\\
e^{-V(u)}  &  =\left(  1+\sin au\right)  ^{2},\nonumber\\
e^{-M}  &  =(\cos au)e^{-V(u)}.\nonumber
\end{align}

Region III:%

\begin{align}
\phi(v) &  =\Gamma(v)=\frac{1}{2}U(v),\\
e^{-U(v)} &  =\cos^{2}bv,\nonumber\\
e^{-V} &  =1,\nonumber\\
\Phi_{0}^{(0)}(v) &  =-b\theta(v),\nonumber\\
A_{y} &  =\sqrt{2}\sin bv,\nonumber\\
e^{-M} &  =\cos bv\nonumber
\end{align}

These initial data imposed from right and left results in the following
interaction region.

Region IV:%

\begin{align}
e^{-U} &  =\cos^{2}au+\cos^{2}bv-1,\\
e^{-V} &  =\left(  1+\sin au\right)  ^{2},\nonumber\\
e^{-M} &  =(\cos au)(\cos bv)e^{-V},\nonumber\\
\phi(u,v) &  =\Gamma(u,v)=\frac{1}{2}U(u,v),\nonumber\\
\Phi_{0}^{(0)} &  =-b\theta(v)(\cos bv)e^{\frac{U}{2}},\nonumber\\
\Phi_{2}^{(0)} &  =\left(  \frac{V_{u}}{2}\right)  \left(  \sin bv\right)
e^{\frac{U}{2}}.\nonumber
\end{align}

It is manifest now, by virtue of the chosen scalar field that we have;%

\begin{align}
\Psi_{2}  &  =0,\\
\Psi_{4}^{(0)}  &  =a\delta(u)-a^{2}\theta(u)\left(  1-\sin au\right)  \left[
\frac{3}{1+\sin au}+2e^{U}\sin au\right]  .\nonumber
\end{align}

We remark that this technique of making $\Psi_{2}=0$, by using a scalar field
may have a larger scope within the context of colliding waves. It may not be
possible however, to find a physical scalar field that satisfies the criteria.
As an example we recall the Khan-Penrose (KP) metric of colliding impulsive
gravitational waves \cite{KP}. Even when we complexify the scalar field we end
up with%

\begin{align}
\sqrt{2}  &  \mid\phi_{u}\mid=\sqrt{\Gamma_{u}U_{u}},\\
\sqrt{2}  &  \mid\phi_{v}\mid=\sqrt{\Gamma_{v}U_{v}}.\nonumber
\end{align}

It can readily be seen from $\Gamma=-M_{KP}$, that $\Gamma_{u}U_{u}<0$ (
$\Gamma_{v}U_{v}<0$\ ) so that such a scalar field does not exist and we can
not make $\Psi_{2}=0$. A similar calculation for the Yurtsever \cite{YU}
solution reveals, however that such a scalar field does exist albeit it is
tedious to be determined.

The fact that the role of a gravitational degree of freedom can be taken over
by a scalar field in the problem of colliding waves is not new. We recall that
by identifying the metric function $V$ by the scalar field $\phi$ the problem
/solution remains invariant \cite{H1}. This amounts to the fact that a
linearly polarized gravitational wave can be imitated by a scalar field $\phi$
with $V=0$. We add that when the gravitational waves are cross polarized such
an identification with either real or complex scalar field fails.

We wish to add that by shifting $v\theta\left(  v\right)  \rightarrow
-v\theta\left(  -v\right)  $ the foregoing solution remains intact while the
physical interpretation ( see Fig(2)) changes drastically \cite{DH}: An
incoming em field coupled with a scalar field characterized by the step
function $\theta\left(  -v\right)  $ reflects from a curved region valid for
$u>0,$ $v<0$ and transforms into an outgoing gravitational field coupled with
scalar wave. This scenario arises as a result of the invariance of the field
equations and their solution under $v\theta\left(  v\right)  \rightarrow
-v\theta\left(  -v\right)  $. Setting the scalar field to zero is interpreted
in anology as an em wave reflecting from the curved region implicated by the
Griffiths's solution. We must admit that the physical interpretation of such a
wave transmutation in general relativity remains still obscure.

\section{Reflection of em Wave From a Succession of Gravitational Impulsive
Waves.}

The generic form of the metric in region II, for parallel polarization can be
taken as%

\begin{equation}
ds^{2}=2dudv-\left(  Fdx\right)  ^{2}-\left(  Gdy\right)  ^{2},
\end{equation}

where $F$ and $G$ are only functions of $u$. For the single impulsive wave
located at $u=0$ we have%

\begin{align}
F  &  =1+u\theta(u),\\
G  &  =1-u\theta(u).\nonumber
\end{align}

Now if we superimpose a second impulsive wave at the wavefront $u=u_{1}>0$,
our $F$ and $G$ functions become%

\begin{align}
F  &  =1+u\theta(u)-\frac{2}{1-u_{1}}\left(  u-u_{1}\right)  \theta\left(
u-u_{1}\right)  ,\\
G  &  =1-u\theta(u)+\frac{2}{1+u_{1}}\left(  u-u_{1}\right)  \theta\left(
u-u_{1}\right)  ,\nonumber
\end{align}

in which the coefficients are chosen deliberately such that%

\begin{equation}
e^{-U}=FG=1-u^{2}\theta(u),
\end{equation}

still holds. Similarly, for the 3-waves case we have%

\begin{align}
F  &  =1+u\theta(u)-\frac{2}{1-u_{1}}\left(  u-u_{1}\right)  \theta\left(
u-u_{1}\right)  +\frac{2}{1-u_{1}}\frac{1+u_{1}}{1+u_{2}}\left(
u-u_{2}\right)  \theta\left(  u-u_{2}\right)  ,\\
G  &  =1-u\theta(u)+\frac{2}{1+u_{1}}\left(  u-u_{1}\right)  \theta\left(
u-u_{1}\right)  -\frac{2}{1+u_{1}}\frac{1-u_{1}}{1-u_{2}}\left(
u-u_{2}\right)  \theta\left(  u-u_{2}\right)  ,\nonumber
\end{align}

which also satisfies (58) for $u_{2}>u_{1}>0$. It is not difficult to write a
similar expression for an arbitrary number of successive waves \cite{H2}. The
remaining metric function $V$ is given by%

\begin{equation}
e^{V}=\frac{F}{G}.
\end{equation}

Since it is the expression $V_{u}$ that enters into the field equations we
wish to give its form for different cases:%

\begin{align}
2-waves  &  :\text{ \ \ }V_{u}=\frac{2}{1-u^{2}}\left[  \theta(u)-2\theta
(u-u_{1})\right]  ,\\
3-waves  &  :\text{ \ \ }V_{u}=\frac{2}{1-u^{2}}\left[  \theta(u)-2\theta
(u-u_{1})+2\theta(u-u_{2})\right]  ,\nonumber\\
n+1-waves  &  :\text{ \ \ }V_{u}=\frac{2}{1-u^{2}}\left[  \theta
(u)+2\sum_{i=1}^{n}\left(  -1\right)  ^{i}\theta(u-u_{i})\right]  .\nonumber
\end{align}

The impulsive wavefronts $u_{i}$ must obviously satisfy $0\leq u_{i}<1$ and we
have the ordering relation $u_{i}<u_{j}$ for $i<j$. In between each of the
successive waves our spacetime is naturally flat. We note that in taking
$\left(  V_{u}\right)  ^{2}$, which appears in the field equations we adopt
\ the standard properties of the step functions, such as%

\begin{equation}
\theta(u-u_{i})\theta(u-u_{j})=\theta(u-u_{j}),
\end{equation}

for $i<j$. As a result we obtain%

\begin{equation}
\left(  V_{u}\right)  ^{2}=\frac{4\theta(u)}{\left(  1-u^{2}\right)  ^{2}},
\end{equation}

which implies that this term effectively is equivalent to a single wave
located at $u=0$. In Fig.3, we show the collision problem of an em wave with a
train of \textit{3-waves }system. The Weyl components of the succesive waves
can be found easily. We have for,%

\begin{align}
2-waves  &  :\text{ \ \ }\Psi_{4}=-\delta(u)+\frac{2}{1-u_{1}^{2}}%
\delta(u-u_{1}),\\
3-waves  &  :\text{ \ \ }\Psi_{4}=-\delta(u)+\frac{2}{1-u_{1}^{2}}%
\delta(u-u_{1})-\frac{2}{1-u_{2}^{2}}\delta(u-u_{2}),\nonumber\\
n+1-waves  &  :\text{ \ \ }\Psi_{4}=-\delta(u)-2\sum_{i=1}^{n}\frac{\left(
-1\right)  ^{i}}{1-u_{i}^{2}}\delta(u-u_{i}).\nonumber
\end{align}

For completeness we remind that for the present impulsive plane wave spacetime
the non-zero Riemann tensor components are related to $\Psi_{4}$ by%

\begin{align}
R_{uxux}  &  =-e^{V-U}\Psi_{4},\\
R_{u%
%TCIMACRO{\unit{y}}%
%BeginExpansion
\operatorname{y}%
%EndExpansion
u%
%TCIMACRO{\unit{y}}%
%BeginExpansion
\operatorname{y}%
%EndExpansion
}  &  =e^{-V-U}\Psi_{4}.\nonumber
\end{align}

For the 2-waves case, as an example, we have%

\begin{equation}
R_{uxux}=\delta\left(  u\right)  -2\frac{1+u_{1}}{1-u_{1}}\delta\left(
u-u_{1}\right)  =-R_{u%
%TCIMACRO{\unit{y}}%
%BeginExpansion
\operatorname{y}%
%EndExpansion
u%
%TCIMACRO{\unit{y}}%
%BeginExpansion
\operatorname{y}%
%EndExpansion
}\nonumber
\end{equation}

It is manifest from these expressions that in between each pair of successive
waves we have a flat space. We remind also that the case of shock
gravitational waves is different. The problem of colliding em wave with a
shock sandwich wave in which the incoming Weyl scalar is of the form%

\begin{equation}
\Psi_{4}=a^{2}\left[  \theta(u)-\theta(u-u_{1})\right]  ,
\end{equation}

was considered before \cite{BAH}. Clearly this case has a uniform curvature
filling the sandwich and is entirely different from the present case. The
solution of colliding em wave (with the same initial data as in the previous
section and without the scalar field) is given by%

\begin{align}
e^{-U} &  =\cos^{2}bv-u^{2},\\
e^{V} &  =\frac{F}{G},\text{ \ \ \ \ \ \ \ }e^{-M}=(\cos bv)\sqrt{1-u^{2}%
}e^{U/2},\nonumber\\
A_{y} &  =\sqrt{2}\left(  \sin bv\right)  e^{-V/2},\nonumber\\
\Phi_{0}^{(0)} &  =-b\theta(v)(\cos bv)e^{\frac{U}{2}}\text{, \ \ \ }\Phi
_{2}^{(0)}=\left(  \frac{V_{u}}{2}\right)  \left(  \sin bv\right)  e^{\frac
{U}{2}},\nonumber
\end{align}

where \ $F$ and $G$ are given in terms of the impulsive waves in succession.
For \textit{2-waves} and \textit{3-waves }\ cases we have given them
explicitly in (57) and (59). It is observed that although the Maxwell
component $\Phi_{0}$ remains unaffected by the successive waves, $\Phi_{2}$
component changes sign each time when the em wave encounters a new wavefront.
This amounts to a phase change by $180^{\circ}$ of the reflected $\Phi_{2}$
component of the em wave. Overall effect is that if it crosses an even number
of impulsive waves we have $\Phi_{2}\rightarrow-\Phi_{2}$ while for an odd
number of waves the sign remains unchanged. This sign change is physically
significant since it shows itself in the em invariant $I=\frac{1}{4}F_{\mu\nu
}F^{\mu\nu}=2\Phi_{0}\Phi_{2}$.

To conclude this section we state that by the \textit{M-shift }method we can
easily extend the present consideration of successive impulsive waves to the
case of Einstein-Maxwell-scalar case, which we shall not elaborate.

\section{Conclusion.}

It is shown that reflection of em waves from gravitational waves with added
scalar fields and with superposition of impulsive waves still satisfies the
solution given long time ago by Griffiths. Solutions were also found in the
past in which gravitational waves reflect from scalar waves while em waves do
not reflect from scalar waves \cite{CH}. In particular we choose our initial
data from the CH forming waves to see whether a non-singular reflection
process results. Our conclusion is that the mutual focusing is still quite
strong to yield a curvature singularity. It remains open to find a CH forming
reflection process yet. We can also exploit the presence of the scalar field
by making a particular choice which sets $\Psi_{2}=0$, giving rise to a novel
case without a Coulomb component in the interaction region. We have given also
a counter example ( the Khan-Penrose case) which implies that this condition
can not be generic.

The second part of the paper presents a multiple reflection process of an em
wave from layers of successive impulsive gravitational waves. While $\Phi_{0}$
remains invariant, \ $\Phi_{2}$ is multiplied by $\left(  -1\right)  ^{n+1}$,
for $n\geq1$, representing the number of superposed waves. This succesive sign
flip in the Maxwell spinor is the precursor of a train of gravitational
impulsive waves and may be searched for an indirect evidence of the latter. As
a final remark we note that the reflection process is applicable only to a
linearly polarized em wave. If cross-polarized em wave is considered, the
condition $V=V(u)$ ( and $W=W(u)$ ) remains no more feasible \cite{H3}. It
remains to be seen whether it works in higher dimensional string theory where
scalar field is replaced by dilaton and em field is replaced by the higher
form fields. From physics standpoint it is an utmost important and fashionable
problem nowadays to understand the classical scattering of wave packets in
general relativity since black holes are believed to be produced in abundance
in such processes \cite{GR}. While investing much effort on reflections from
the horizons of black holes, it will not be adequate to ignore reflection
processes of waves from other systems, as considered in this paper.\newline

\begin{acknowledgments}
One of us (O.G) would like to thank to the Physics Department of the Bogazici
University for their kind hospitality. 
\end{acknowledgments}

\begin{center}
FIGURE CAPTIONS
\end{center}

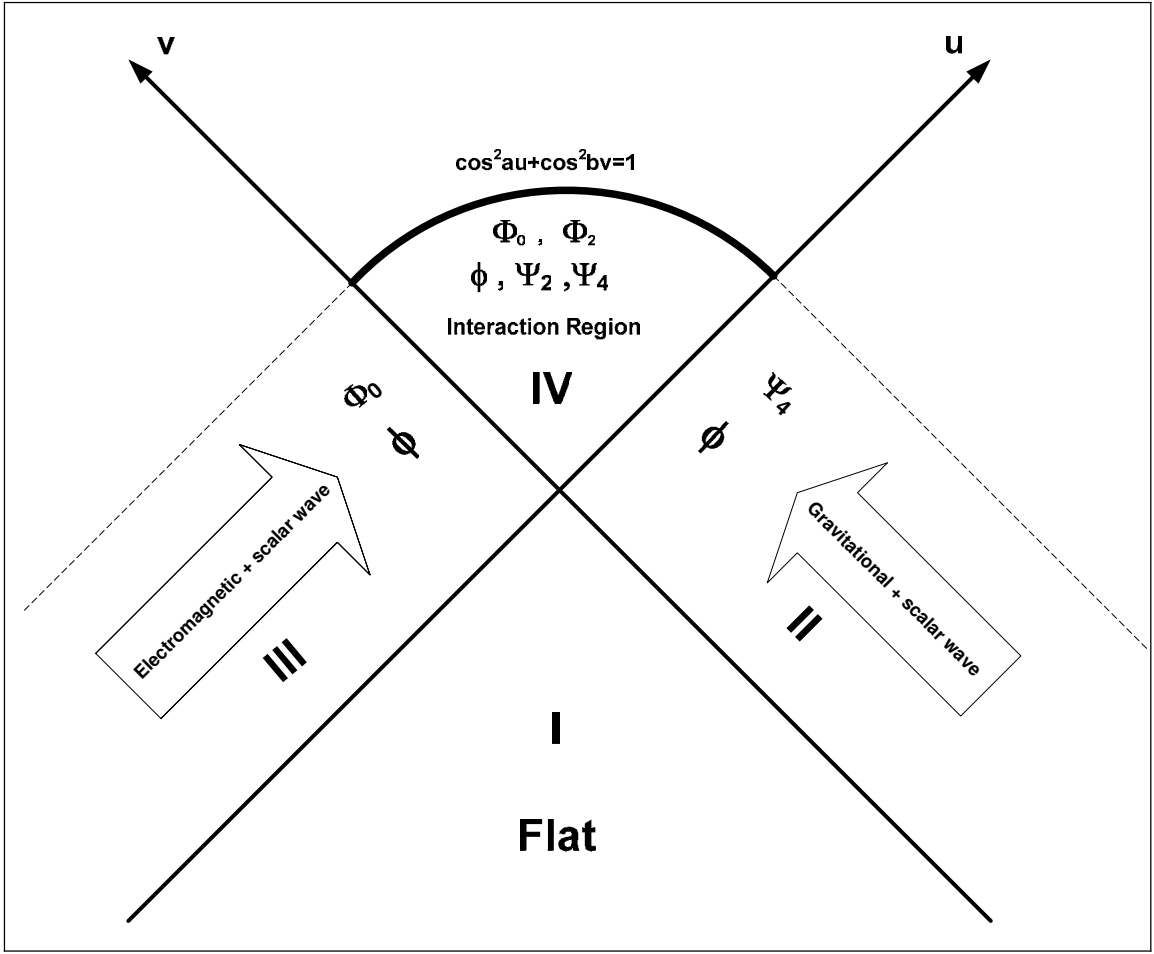
Figure 1: The space-time diagram describes the collision of wave packets that
contains em + scalar waves in one of the incoming region and gravitational +
scalar waves in the other incoming region. In this process the em wave still
partly reflects.

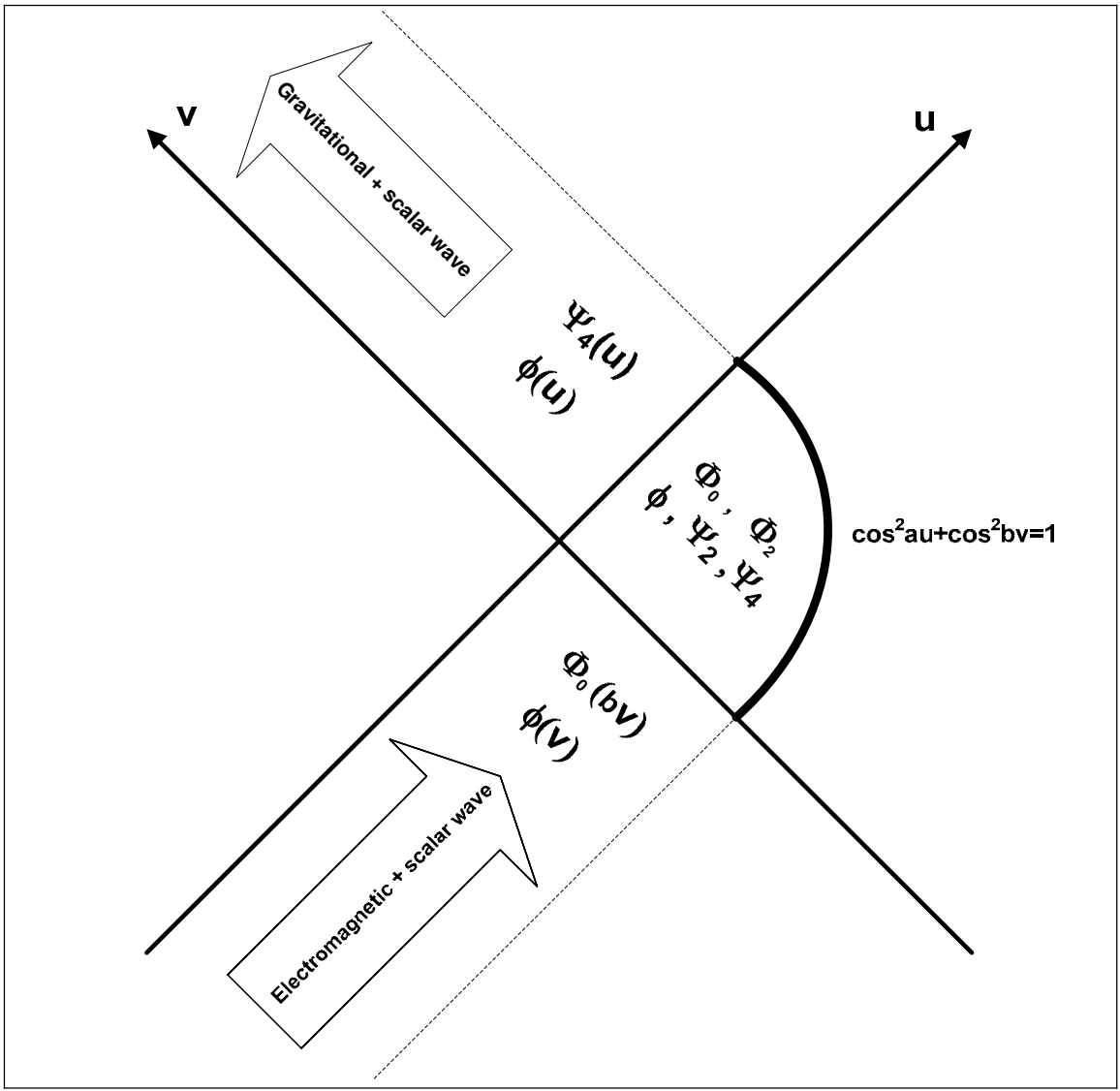
Figure 2: The space-time diagram describes, reflection of an em + scalar wave
packet from a curved spacetime with a timelike singularity and transforming
into an outgoing gravitational + scalar wave packet. This kind of
interpretation is the result of applying $v\theta(v)$ $\rightarrow$
$-v\theta(-v)$ in figure 1.

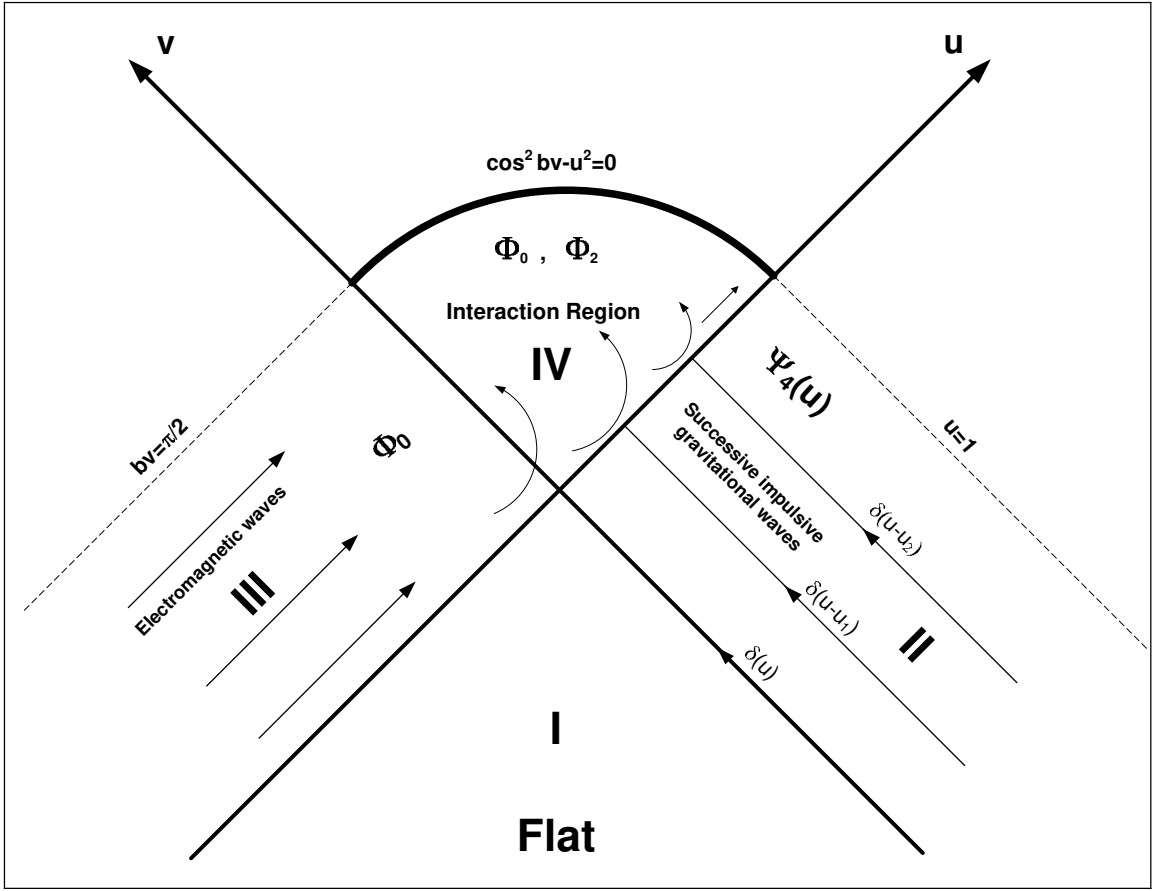
Figure 3: The space-time diagram describes the reflection process of an em
wave from a succession of three impulsive gravitational waves. The process
results in a curvature singularity on the focusing surface $\cos^{2}%
bv-u^{2}=0.$

\end{document}